\documentclass[lettersize,journal]{IEEEtran}
\usepackage{cite}
\usepackage{amsmath,amssymb,amsfonts,amsthm,mathrsfs}
\usepackage{multirow}
\usepackage{bbm}
\usepackage{algorithmic}
\usepackage{caption}
\usepackage{graphicx}
\usepackage{tikz} 
\usetikzlibrary{automata, positioning, arrows}
\usepackage{textcomp}
\usepackage{xcolor}
\usepackage{manfnt}
\usepackage{mathtools}
\usepackage{fixmath}
\usepackage{pgf}
\usepackage{xcolor}
\usetikzlibrary{positioning,arrows}
\usepackage{balance} 

\def\BibTeX{{\rm B\kern-.05em{\sc i\kern-.025em b}\kern-.08em
    T\kern-.1667em\lower.7ex\hbox{E}\kern-.125emX}}
    
\newcommand{\N}{\mathbb{N}}

\newcommand{\R}{\mathbb{R}}
\newcommand{\Q}{\mathbb{Q}}

\newcommand{\sC}{\mathcal{C}}
\newcommand{\sD}{\mathcal{D}}
\newcommand{\sF}{\mathcal{F}}

\newcommand{\sT}{\mathcal{T}}

\newcommand{\sX}{\mathcal{X}}

\theoremstyle{plain}
\newtheorem{thm}{Theorem}
\newtheorem{lem}{Lemma}

\theoremstyle{defn}
\newtheorem{defn}{Definition}

\theoremstyle{rem}
\newtheorem{rem}{Remark}

\tikzstyle{block} = [draw, rectangle, 
  minimum height=3em, minimum width=4em]

\newcommand{\tcr}[1]{\textcolor{red}{#1}}

\newcommand{\tco}[1]{\textcolor{orange}{#1}}

\renewcommand{\tcr}[1]{#1}
\renewcommand{\tco}[1]{#1}

\begin{document}

\title{Characterization of the Complexity of Computing the Capacity of Colored Noise Gaussian Channels}

\author{Holger Boche, Andrea Grigorescu, Rafael F. Schaefer, and H. Vincent Poor}
\maketitle

\begin{abstract}
This paper explores the computational complexity involved in determining the capacity of the band-limited additive colored Gaussian noise (ACGN) channel and its capacity-achieving power spectral density (p.s.d.). The study reveals that when the noise p.s.d. is a strictly positive computable continuous function, computing the capacity of the band-limited ACGN channel becomes a $\#\mathrm{P}_1$-complete problem within the set of polynomial time computable noise p.s.d.s. Meaning that it is even more complex than problems that are $\mathrm{NP}_1$-complete. Additionally, it is shown that the capacity-achieving distribution is also $\#\mathrm{P}_1$-complete.
Furthermore, under the widely accepted assumption that $\mathrm{FP}_1 \neq \#\mathrm{P}_1$, it has two significant implications for the ACGN channel. The first implication is the existence of a polynomial time computable noise p.s.d. for which the computation of its capacity cannot be performed in polynomial time, i.e., the number of computational steps on a Turing Machine grows faster than all polynomials. The second one is the existence of a polynomial time computable noise p.s.d. for which determining its capacity-achieving p.s.d. cannot be done within polynomial time.
\end{abstract}

\footnotetext[1]{This work of H. Boche was supported in part by the German Federal Ministry of Education and Research (BMBF) within the national initiative on 6G Communication Systems through the research hub 6G-life under Grant 16KISK002, within the national initiative on Post Shannon Communication
(NewCom) under Grant 16KIS1003K, and the project Hardware Platforms and Computing Models for Neuromorphic Computing (NeuroCM) under Grant 16ME0442. It has further received funding by the German Research Foundation (DFG) within Germany’s Excellence Strategy EXC-2092 – 390781972.

 This work of R. F. Schaefer was supported in part by the BMBF within NewCom under Grant 16KIS1004, and within 6G-life under Grant 16KISK001K as well as in part by the German Research Foundation (DFG) as part of Germany’s Excellence Strategy – EXC 2050/1 – Project ID 390696704 – Cluster of Excellence “Centre for Tactile Internet with Human-in-the-Loop” (CeTI) of Technische Universität Dresden and under Grant SCHA 1944/11-1.
 
 This work of H. V. Poor was supported by the U.S National Science Foundation under Grant CNS-2128448.}
  
\footnotetext[2]{H. Boche is with the Chair of Theoretical Information Technology, Technical University of Munich, Arcisstr. 21, 80333 M\"unchen, Germany. (E-mail:boche@tum.de). He is also with the BMBF Research Hub 6G-life and with the Excellence Cluster Cyber Security in the Age of Large-Scale Adversaries (CASA), Ruhr University Bochum.
A. Grigorescu is with the Chair of Theoretical Information Technology, Technical University of Munich, Arcisstr. 21, 80333 M\"unchen, Germany. (E-mail:andrea.grigorescu@tum.de).
R. F. Schaefer is with the Chair of Information Theory and Machine Learning, Technische Universit\"at Dresden, the BMBF Research Hub 6G-life, the Cluster of Excellence ``Centre for Tactile Internet with Human-in-the-Loop (CeTI)'', and the 5G Lab Germany, Technical University of Dresden, 01062 Dresden, Germany (E-mail: rafael.schaefer@tu-dresden.de).
H.V. Poor is with the Department of Electrical and Computer Engineering, Princeton University (E-mail:poor@princeton.edu)}

\section{Introduction}



The band-limited additive colored Gaussian noise (ACGN) channel plays a crucial role in representing commonly encountered channels in wireless communication, such as the frequency selective fading channel. The band-limited Gaussian channel, initially introduced in the works of \cite{shannon1949communication, shannon1949communicationnoise}, is a channel that operates in continuous time. In \cite{shannon1949communicationnoise}, Shannon also introduced two noise models in their study: white Gaussian noise and colored Gaussian noise. Colored Gaussian noise follows a Gaussian distribution and has a power spectral density (p.s.d.) that varies with frequency, while white Gaussian noise has a constant spectral density across all frequencies. The capacity and error performance of codes designed for the band-limited ACGN channel were carefully studied in \cite{wyner1966capacity, ash1963capacity, shannon1959probability}. A detailed description of the band-limited ACGN channel and its results can be found in \cite{ash2012information,cover1999elements,ihara1993information,tse2005fundamentals,pinsker1964information}. In \cite{gallager1968information}, Gallager showed that the capacity-achieving p.s.d. of the linear ACGN channel can be determined using the water pouring technique. In \cite{forney1998modulation}, the authors provide an overview of techniques desired to construct capacity-achieving codes for the ACGN channel. 

Computing the capacity of the band-limited ACGN channel is a very important task for the design of practical systems, especially for 6G. The capacity serves as a benchmark for designing and optimizing systems to achieve optimal message transmission. This will enable the construction of codes that fulfill the anticipated reliability and efficiency requirements of 6G systems while also optimizing the use of communication resources. For instance, resource allocation will most likely take place at the base stations of 6G systems. At present and in the foreseeable future, only digital hardware can be used for this task. 

 To address algorithmic computability, we use the concept of a \emph{Turing Machine} (TM) \cite{turing1936computable,turing1938computable,weihrauch2000computable}, which is a mathematical model of an abstract machine that manipulates symbols on a strip of tape according to certain given rules. Any algorithm can be translated into a sequence of steps that can be executed by a TM and therefore provides a simple and very powerful model of computation. TMs have no limitations on computational complexity, computing capacity or storage, and execute programs completely error-free. Accordingly, they provide fundamental performance limits for today's digital computers. TMs account for all those problems and tasks that are algorithmically computable on a classical (i.e., non-quantum) machine.

In \cite{boche2023algorithmic}, it was shown that there exist band-limited computable ACGN channels whose capacities yield non-computable numbers. This implies that the capacity $C$ of the band-limited ACGN channel is in general a non-computable function of its parameter, in this case, the noise p.s.d. $N$ and power constraint $P$. The question now is whether it is possible to restrict the set of band-limited ACGN channels such that the capacity becomes computable for this set of channels. In other words, is there a TM that computes an approximation $\tilde{C}$ of the capacity $C(N,P)$? This means, that for every precision $M\in\N$, the TM computes an approximation $\tilde{C}$ such that
\begin{equation*}
	|\tilde{C}-C(N,P)|\leq \frac{1}{2^M},
\end{equation*}
i.e., the TM stops whenever the desired precision is achieved.

In this paper, our objective is to establish a sufficient condition for an ACGN channel to have a computable capacity. We demonstrate that the capacity of band-limited ACGN channels can be computed if the noise p.s.d. is a strictly positive, computable function of the frequency. Furthermore, we focus on exploring the computational complexity associated with approximating the capacity of these channels. Specifically, we investigate the computational complexity of approximating the channel capacity when the input parameters, namely $N$ and $P$, exhibit low complexity.

In the field of information and communication theory, various problems have been extensively investigated in terms of their computational complexity. The main focus of these studies is to determine whether a given computational problem falls under the $\mathrm{P}$ or $\mathrm{NP}$ complexity class.
The decision problem of determining whether the zero-error capacity and Holevo capacity for quantum channels exceed a certain value $k \in \N$ was proven to be NP-complete \cite{beigi2007complexity}. This result also holds true for the zero-error capacity of discrete memoryless channels \cite{shannon1956zero, sipser1996introduction}. In \cite{langberg2011hardness}, it was demonstrated that deciding whether the network coding capacity is greater or egual to $1$ is an $\mathrm{NP}$-complete problem. The syndrome decoding problem for the Hamming distance was shown to be NP-complete in \cite{berlekamp1978inherent}.
Later, in \cite{vardy1997intractability, vardy1997algorithmic}, it was established that computing the minimum distance of a binary linear code is $\mathrm{NP}$-hard, while the corresponding decision problem is $\mathrm{NP}$-complete. The complexity of decoding minimum distance problems for rank codes within the $\mathrm{NP}$ complexity class is investigated in \cite{gaborit2016hardness}. In \cite{ordentlich2011two}, it was demonstrated that two-dimensional maximum-likelihood sequence detection is $\mathrm{NP}$-hard. The complexity of the capacity region of a classical multiple access channel (MAC) was examined in \cite{leditzky2020playing}. By combining the findings from \cite{leditzky2020playing} and \cite{calvo2010computation}, it was established that determining whether a rate point belongs to the MAC capacity region is an $\mathrm{NP}$-hard problem. The study conducted in \cite{mondelli2014polar} focused on approximation methods for finite blocklength coding as well as $\mathrm{NP}$-complete problems. Additionally, the performance characteristics of suboptimal decoding algorithms were investigated, utilizing the results presented in \cite{eslami2013finite}. 
Resource allocation and scheduling for wireless communication systems lead to $\mathrm{NP}$-complete problems in a wide variety of scenarios, and as a result, there is a vast body of work dedicated to studying these NP-complete resource allocation problems and their approximability, see \cite{sun2018distributed,bagaa2017optimal,chattopadhyay2017gibbsian,zhang2019cellular}.

Our findings reveal that when considering polynomial time strictly positive noise p.s.d.s, the capacity of band-limited ACGN channels becomes computable. Moreover, we show that the computation of $C(N,P)$ as a specific numerical value belongs to the $\#\mathrm{P}_1$ complexity class. The class $\#\mathrm{P}_1$ extends the class $\mathrm{NP}$ to the counting problem. It consists of problems that count the number of solutions of a problem that can be verified by a TM in polynomial time. Additionally, we delve into the complexity of computing the p.s.d. that achieves the channel capacity for ACGN channels. Our findings reveal that both the approximation of capacity and the computation of the capacity-achieving p.s.d. are $\#\mathrm{P}_1$-complete. Therefore, if the assumption $\mathrm{FP}_1\neq\#\mathrm{P}_1$ holds true, then the computation of capacity and the capacity-achieving p.s.d. exhibit high complexity outputs when the input has low complexity. This behavior indicates a complexity blowup, emphasizing the intricate nature of these computations.

The remainder of the paper is organized as follows: In Section \ref{sec:Gauss_Channel}, we introduce the band-limited ACGN channel and present its capacity results. In Section \ref{sec:PF}, we describe the problem formulation. In Section \ref{sec:computability_numbers}, we introduce the computability framework, complexity framework, and classes. In Section \ref{sec:BlowupACGN}, we demonstrate that the capacity of ACGN channels and the capacity-achieving distribution show a complexity blowup behavior.  To support our claims, we present some interim results in the appendix. Finally, we provide our conclusions in Section \ref{sec:conclusion}. 
\section{Continuous Gaussian Channels}
\label{sec:Gauss_Channel}
In this section we consider a communication scenario where both the input and output of the channel are amplitude- and time-continuous. Amplitude-continuous means that the signal alphabets are uncountably infinite, and by time-continuous we allow the transmission to be continuous over time.
The time continuous additive Gaussian channel is represented by the following formula:
\begin{equation*}
	y(t)=x(t)+ z(t), 
\end{equation*}
where $x(t), y(t)$, and $z(t)$ are the channel input, channel output, and noise at time instant $t\in\sT\subset\R$ and they take values in $\R$. The noise $z$ is zero-mean Gaussian distributed. The Fourier \tco{transform} of the input signal is represented by $X=\sF\{x\}$, where $\sF$ is the Fourier operator.

Let $x_T$ be the fraction of the signal $x$ that is equal to $x(t)$ for $t\in[-\frac{T}{2}, \frac{T}{2}]$ and $0$ outside. The total signal power $P_{\text{tot}}$ is given by 
\begin{align*}
P_{\text{tot}}&=\lim_{T\rightarrow\infty}\frac{1}{T}\int_{-T}^{T}|x_T(t)|^2\,dt=\lim_{T\rightarrow\infty}\frac{1}{T}\int_{-T}^{T}\sF|x_T(t)|^2\,df\\
&=\int_{-\infty}^{\infty}P_x(f)\,df
\end{align*} 
where $P_x(f)= \lim_{T\rightarrow\infty}\frac{\sF|x_T(t)|^2}{T}$ is the p.s.d. of $x$.

Similarly, $N(f)$ represents the p.s.d. of the noise $z$.

We consider only band-limited signals. Letting the bandwidth be $B>0$, the p.s.d. of band-limited signals with bandwidth $B$ has the following structure: 
\begin{equation*}
	P_x(f)=\begin{cases}
	P_x(f) &\text{for } f\in [0,B]\\
		0 &\text{else}.
	\end{cases}
\end{equation*}

$z$ is a band-limited colored Gaussian noise with a p.s.d. with the following structure: 
\begin{equation*}
	N(f) = \begin{cases}
		\geq 0 &\text{for } f\in [0,B]\\
		0 & \text{else}.
	\end{cases}
\end{equation*}

We consider a communication scenario subject to a power constraint $P$. This means that the total signal power should not exceed $P$, and it is described by 
\begin{equation*}
	\int_{0}^B P_x(f)\,df\leq P.
\end{equation*}

We consider the set $\sX(B,T,P)$ which is the set of approximately band-limited signals with bandwidth $B$, approximately time-limited to $T$ seconds and with a total power not exceeding $P$, i.e., for every signal $x\in\sX(B,T,P)$ it holds that $\int_{0}^B P_x(f)\,df\leq P$. The goal is to find codes, with codewords in $\sX(B,T,P)$, such that one can reliably distinguish $M$ different signal functions of duration $T$, i.e., that can operate reliably at positive transmission rate $R=\frac{1}{T}\log M$. The capacity $C$ of the band-limited ACGN channel is the maximal rate, for which one can find codes that operate reliably. The capacity of ACGN channels was derived by Shannon in \cite{shannon1949communicationnoise}.

\begin{thm}[\cite{shannon1949communicationnoise}]\label{thm:Shannon}
	\tcr{The capacity of the band-limited ACGN channel with bandwidth $B$, and continuous noise p.s.d. $N$ on the interval $[0,B]$ subject to a power constraint $P>0$ is given by }
	\begin{equation*}
		C(N,P)=\int_0^B \ln\Big(1+\frac{P_x^*(f)}{N(f)}\Big)\,df.
	\end{equation*}
The capacity-achieving p.s.d. is unique and is given by
	\begin{equation}
	P_x^*(f)=\begin{cases}
	\Big[L - N(f)\Big]_+ &\text{ for } f\in[0,B]\\
	0 &\text{f} \notin [0,B],
	\end{cases}
	\end{equation}
	where $L$ is chosen such that $\int_0^B P_x^*(f)\,df=P$ is satisfied.
\end{thm} 
Shannon himself validated Theorem \ref{thm:Shannon} by approximating the channel through a discrete transmission problem in \cite{shannon1949communicationnoise}. To do this, he used tools from convex optimization and essentially followed the water pouring argumentation.

 Water pouring is well known for its simple derivation \cite{flanagan2006proving}.
 In general, the problem is approached by dividing the noise spectrum into $n$ sub-channels of width $\Delta f_n$ and assuming that each sub-channel is \tco{independent} of the others. $N(f)$ is then approximated by $N(f_i)$ for $f\in[f_i-\frac{\Delta f_n}{2},f_i+\frac{\Delta f_n}{2} ]$ and $i\in\{1,\dots, n\}$. The capacity of each sub-channel $f_i$ is given by
 	\begin{equation*}
 	C_n(N,P,f_i)=\Delta f_n \ln \Big(1+\frac{P_x^*(f_i)}{N(f_i)}\Big)
 	\end{equation*}
where $P_x^*(f_i) =L-N(f_i)$ and $L$ is derived by using the method of Lagrange multipliers. The total capacity and the total transmit power are given by 
 	\begin{align}
 	C_n(N,P)&= \sum_{i=1}^n \Delta f_n \ln \Big(1+\frac{P_x^*(f_i)}{N(f_i)}\Big)\label{eq:cn}\\
 	P &= \sum_{i=1}^n \Delta f_n P_x^*(f_i).\label{eq:pn}
 	\end{align}
 As $n\rightarrow\infty$ then $\Delta f_n\rightarrow 0 $ and Eqs. \eqref{eq:cn} and \eqref{eq:pn} become integrals: 
 	\begin{align}
 	C(N,P)&=\lim_{n\rightarrow \infty}C_n(N,P)= \int_0^B \ln\Big(1+\frac{P_x^*(f)}{N(f)}\Big)\,df\label{eq:c_apprx}\\
 	P &= \lim_{n\rightarrow \infty}\sum_{i=1}^n \Delta f_n P_x^*(f_i)=\int_0^B P_x^*(f)\,df.\label{eq:p_apprx}
 	\end{align}

Note that Shannon in \cite{shannon1949communicationnoise} and subsequent literature, such as \cite{gallager1968information,flanagan2006proving}, did not specify an effective estimation for the convergence speed of \eqref{eq:c_apprx} and \eqref{eq:p_apprx}. On one hand, the solution of the discrete optimization problem is easy to obtain. On the other hand, this solution does not provide any insights into the computational complexity involved in calculating the capacity and the capacity-achieving p.s.d. of the continuous problem. These questions are addressed in Section \ref{sec:PF} and Section \ref{sec:BlowupACGN}.

\section{Problem Formulation}
\label{sec:PF}

We consider the band-limited ACGN channel with noise p.s.d. $N$ on $[0,B]$. $N$ is strictly positive. For the average noise power spectrum $\bar{N}$ with	
\begin{equation*}
	\bar{N}=\frac{1}{B}\int_{0}^B N(f)\,df
\end{equation*}
and a power constraint $\mathrm{P}$ with $P>\bar{N}$, from Theorem \ref{thm:Shannon} we have that the the signal's optimal p.s.d. is given by $P_x^*(P,f)=(L-N(f))$ with $\int_0^{B}P_x^*(P,f)\,df=P$. That is
\begin{equation*}
LB-\int_{0}^BN(f)\,df =P,
\end{equation*}
which implies that 
\begin{equation*}
L=\frac{P}{B}+ \frac{1}{B}\int_{0}^BN(f)\,df.
\end{equation*}
From Theorem \ref{thm:Shannon} we have that the capacity of the band-limited ACGN channel with power constraint $\mathrm{P}$ and noise p.s.d. $N$ is 
\begin{align}
	C(P,N)&=\int_{0}^B\ln(P_x^*(P,f)-N(f))\,df-\int_{0}^B\ln(N(f))\,df\nonumber\\
	&= B\ln L - \int_0^B\ln(N(f))\,df.\label{eq:cL}
\end{align}
We consider noise spectral densities $N$ which are infinitely differentiable, strictly positive and can be computed in polynomial time. For these channels, we are interested in the capacity-achieving p.s.d.s $P_x^*$ whose values are computable numbers. This prompts the following question:

\emph{Question 1}: What is the computational complexity of computing the values of the capacity-achieving p.s.d. $P_x^*(P,\cdot)$ as a function of the frequency?

In \cite{boche2023algorithmic}, it was shown, that there are ACGN channels with computable continuous non-negative p.s.d.s, whose capacities yield a non-computable number. 
We consider noise spectral densities $N$ which are infinitely differentiable, strictly positive and can be computed in polynomial time.
Assume $\mathrm{P}$ is a rational number. We consider the computable subset of such $N$ and $\mathrm{P}$ whose capacity yields a computable number. Namely, let $M$ be the desired precision for the computation of the capacity, i.e., the capacity lies at most $\frac{1}{2^M}$ away from the computed approximation. Then there exists a TM $M_C$, that takes $P,N,M$ as input and computes an approximation $\tilde{C}= M_C(P,N,M)$ of the capacity $C(P,N)$ of the ACGN channel with p.s.d. $N$ and power constraint $\mathrm{P}$ such that $|\tilde{C}-C(P,N)|\leq \frac{1}{2^M}$.
We are interested in knowing how much time does such a TM need to compute the capacity approximation within the desired precision. This prompts the following question:

\emph{Question 2}: What is the computational complexity of computing the capacity $C(P,N)$?

We formalize and answer Question 1 and Question 2 in Section \ref{sec:BlowupACGN} using the framework of complexity theory introduced in Section \ref{sec:computability_numbers}.

Currently, the computation of problems is limited to digital machines. Therefore, to examine the fundamental limits of today's computers and address the former questions, the application of Alan Turing's computability theory, specifically TMs, is of central importance.

\section{Computability and Complexity Framework}
\label{sec:computability_numbers}
In this section, we introduce the fundamental concepts of computability and complexity theory, which are needed therefore. Computability and computable real numbers were initially proposed by Turing in \cite{turing1936computable} and \cite{turing1938computable}. In this context, computable numbers denote real numbers that can be computed using TMs.

\subsection{Computability}
To define computable numbers and determine the time complexity of its computation we first need to study its representation. For this, we introduce the concept of dyadic rational numbers.

The set $\sD$ of \emph{dyadic rational numbers} are rational numbers with finite binary expansion. Each dyadic rational number $d$ is naturally represented by a binary string $s=\pm s_ns_{n-1}\cdots s_0.t_1t_2\cdots t_m$ satisfying
\begin{equation*}
	d=\pm\sum_{i=1}^ns_i2^i\pm\sum_{j=1}^mt_j2^{-j}.
\end{equation*}
A representation $s$ of a dyadic rational $d$ has the precision $m$ ($\text{prec}(s)=m$), if it has $m$ bits to the right of the binary point.

\begin{defn}\label{def:computable number}
A number $t\in\R$ is said to be computable if there exists a TM $M$ with input $n\in\N$ and output $\varphi(n)=M(n)\in\Q$ such that 
\begin{equation}\label{eq:binaryconv}
	|t-\varphi(n)|\leq 2^{-n}.
\end{equation}
We say that $\varphi(n)$ binary converges to $t$ if \eqref{eq:binaryconv} is satisfied. The set $\R_c\subset\R$ is the set of all computable numbers. 
\end{defn}

A \emph{function-oracle TM} is an ordinary TM $M$ equipped with an additional query tape and two additional states: the query state and the answer state. When the machine enters the query state, the oracle function $\varphi$ replaces the current string $s$ in the query tape by the string $\varphi(s)$, moves the tape head back to the first cell of the query tape and puts the machine $M$ in the answer state. When the time complexity is considered, the entire process of querying for the value $\varphi(s)$ costs only one time unit to the machine. 

Next we introduce the notion of computable functions using the concept of oracle TMs.

\begin{defn}
A real function $f\colon\R\rightarrow\R$ is \emph{computable} if there is a function-oracle TM $M$ such that for each $x\in\R$ and each $\varphi$ that binary converges to $x$, the function $\psi$ computed by $M$ with oracle $\varphi$ (i.e., $\psi(n)=M_{\varphi}(n)$) binary converges to $f(x)$.
\end{defn}

Intuitively, a function $f$ is computable, if for a given $x$ and the oracle $\varphi$, the oracle TM $M_\varphi$ takes $n$ as input and computes a dyadic rational $M_{\varphi}(n)$ that binary converges to $f(x)$, i.e., $|M_{\varphi}(n)-f(x)|\leq 2^{-n}$. During the computation, the information about $x$ can be obtained from the oracle $\varphi$ in one time step.

An important property of a computable real function is that it must be continuous on its domain and the modulus of continuity is computable.

\begin{defn}
	Let $f\colon[a,b]\rightarrow\R$ be a continuous function on $[a,b]$. Then, a function $m\colon\N\rightarrow\N$ is said to be a \emph{modulus function} of $f$ on $[a,b]$, if $|x-y|\leq 2^{-m(n)}$ implies $|f(x)-f(y)|\leq 2^{-n}$, for all $x,y\in[a,b]$ and for all $n\geq0$.
\end{defn}
	
\begin{thm}
 If $f\colon [a, b]\to \R$ is computable on $[a, b]$ then $f$ is continuous on $[a, b]$; furthermore, $f$ has a recursive modulus function $m$ on $[a, b]$.
\end{thm}	
We denote the set of all computable continuous functions on $[a,b]$ by $\sC_c([a,b])$.
\subsection{Complexity of Real Functions}

For any $t\in\R_c$, the TM will need several iterations to compute $\varphi(n)$. The number of iterations needed will increase when $n$ increases. The quantitative relation between the number of iterations for computing $\varphi(n)$ and $n$ determines the computational complexity of $t\in\R_c$.

\begin{defn}
 Let $t$ be an integer function. The \emph{time complexity} of a computable real number $x$ is bounded by $t$ if there exists a TM which computes, on each input $n\in\N$, a dyadic rational number $d$ in $t(n)$ moves such that $|d-x|\leq2^{-n}$.
\end{defn}

\begin{defn}
A real number $x$ is \emph{polynomial time computable} if its time complexity is bounded by a polynomial function $\mathrm{P}$.
\end{defn}

\begin{defn}
Let $f\colon[0,1]\rightarrow\R$ be a computable function. The \emph{complexity} of $f$ on $[0,1]$ is bounded by a function $q\colon\N\rightarrow\N$ if there exists an oracle TM $M$ which computes $f$ such that for all $\varphi$ that binary converge to a real number $x\in[0,1]$ and for all $n>0$, $M_{\varphi}(n)$ halts in at most time $q(n)$.
\end{defn}

\begin{defn}
A real function $f\colon[0,1]\rightarrow\R$ is \emph{polynomial time computable} if its time complexity is bounded by a polynomial function $\mathrm{P}$.
\end{defn}
%

\subsection{Complexity Classes}
In this subsection, we introduce some complexity classes that characterize the complexity of solving certain problems. We introduce and discuss the complexity classes related to
decision and counting problems.

For decision problems, the best known complexity classes are $\mathrm{P}$ and $\mathrm{NP}$. These are problems that for a given input string only have two possible solutions, ``$0$'' = ``no” or ``$1$” = ``yes”.

Consider the language $\Sigma \subset \{0,1\}^*$\footnote{\emph{Notation:} $\{0,1\}^*$ denotes the set of all finite words in $\{0,1\}$, $|x|$ denotes the length of the sequence $x$.}
which is an infinite subset. The Entscheidungsproblem is described as follows: Find a TM $M$, such that for $x\in\{0,1\}^*$ it holds that $M(x)=1$ if and only if $x\in\Sigma$, otherwise it holds that $M(x)=0$, i.e., the output of $TM$ is either $0$ or $1$.
The class $\mathrm{P}$ is the set of all decision problems that can be solved by a deterministic TM in polynomial time, i.e., in a computation time that grows polynomial in the input size.
The class $\mathrm{NP}$ is the set of all problems that can be solved by a non-deterministic TM in polynomial time. 
 It is clear from the definition that $\mathrm{P} \subset \mathrm{NP}$ but it remains an open question whether $\mathrm{P} = \mathrm{NP}$ or $\mathrm{P} \varsubsetneq \mathrm{NP}$. It is widely assumed that $\mathrm{P}$ is a proper subset of $\mathrm{NP}$.

In order to formally define the complexity classes, we consider the concept non-deterministic TMs(NDTMs). The difference between a deterministic TM and a NDTM is that the latter has more than one possible move from a given configuration, and also a special state $q_{accept}$. Hence, for a NDTM $M$, $M$ outputs $1$ on a given input $x$ if there is at least one sequence of moves for $x$ that makes $M$ reach $q_{accept}$, otherwise, if $x$ makes $M$ stop, then it reaches the halting state and $M(x)=0$. A NDTM is polynomial, if there is a $p\colon\N\rightarrow\N$ such that for every $x$ the NDTM reaches the halting state or $q_{accept}$ in at most $p(|x|)$ steps. 

\begin{defn}[Classes $\mathrm{P}$ and $\mathrm{NP}$ ]\label{def:PNP}
Let $\Sigma\subset\{0,1\}^*$ be a language. Then $\Sigma$ is in \emph{$\mathrm{P}$ } if there exists a TM $M$ that solvers the Entscheidungsproblem for $\Sigma$ in polynomial time.

A language $\Sigma\subset\{0,1\}^*$ is in \emph{$\mathrm{NP}$ }, if there exists a polynomial time NDTM $M$ such that $x\in\Sigma\Leftrightarrow M(x)=1.$
\end{defn}
The name $\mathrm{NP}$ comes originally from \textbf{N}on-deterministic \textbf{P}olynomial TMs. Intuitively, the class $\mathrm{NP}$ is the class of all problems, which are verifiable in polynomial time. This can be ilustrated in the following way: Let $M$ be a polynomial NDTM. Let $y$ be a sequence of moves describing a non-deterministic path that makes $M$ reach $q_{accept}$ on input $x$, then $y$ is a \emph{certificate} for $x$. This certificate has length $p(|x|)$ and can be then verified by a polynomial time TM, which checks that $M$ would have entered $q_{accept}$ after using the non-deterministic path of $M$.
In other words, if a set $\Sigma\subset\{0,1\}^*$ is in $\mathrm{NP}$, then there exists a TM $M_v$ that takes inputs from $\{0,1\}^*\times \{0,1\}^*$ and outputs values in $\{0,1\}$ such that for any $x\in\Sigma$, there exists a certificate $y\in\{0,1\}^{p(|x|)}$ satisfying $M_v(x,y) = 1$. 

Next, we consider functions that are defined on a different domain. More precisely, we consider a complexity class similar to $\mathrm{P}$ and $\mathrm{NP}$ that contains functions whose input belong to a singleton alphabet, i.e., functions whose domain is $\{0\}^*$.

\begin{defn}[Classes $\mathrm{P}_1$ and $\mathrm{NP}_1$]\label{def:PNP1}
Let $\Sigma_1\subset\{0\}^*$ be a language. Then $\Sigma_1$ is in \emph{$\mathrm{P}_1$ } if there exists a TM $M$ that solvers the Entscheidungsproblem for $\Sigma_1$ in polynomial time. 
A language $\Sigma_1\subset\{0\}^*$ is in \emph{$\mathrm{NP}_1$ }, if there exists a non-deterministic polynomial time TM $M$ such that $x\in\Sigma_1\Leftrightarrow M(x)=1.$
\end{defn}
%
%
Let us introduce the characteristic functions $\chi_{\Sigma}\colon\{0,1\}^*\rightarrow\{0,1\}$ and $\chi_{\Sigma_1}\colon\{0\}^*\rightarrow\{0,1\}$ for languages $\Sigma\subset\{0,1\}^*$ and $\Sigma_1\subset\{0\}^*$, respectively. These characteristic functions evaluate as follows: $\chi_{\Sigma}(x)=1 \Leftrightarrow x\in\Sigma$.

We can then consider Definitions \ref{def:PNP1} and \ref{def:PNP} as complexity requirements for computing the characteristic function $\chi_{\Sigma}$. In other words, we can determine whether $\Sigma$ is in the complexity class $\mathrm{P}$ and whether $\Sigma_1$ is in the complexity class $\mathrm{P}_1$ by investigating whether the characteristic function $\chi_{\Sigma}$ is polynomial time computable.

We want to illustrate the same notions for the complexity classes for general functions.
 To represent such problems, we use functions denoted as $f: \{0, 1\}^* \rightarrow \N$, defined on the set of all finite words in the binary alphabet $\{0, 1\}$. For any given word $x \in \{0, 1\}^*$, the value of $f(x)$ represents the count of solutions for that particular instance. The classes analog to the classes $\mathrm{P}$ and $\mathrm{NP}$ for counting problems are denoted by $\mathrm{FP}$ and $\#\mathrm{P}$ respectively.

\begin{defn}[Classes $\mathrm{FP}$ and $\#\mathrm{P}$]\label{def:FPsharpP}
A function $f \colon \{0, 1\}^* \rightarrow \N$ is in $\mathrm{FP}$ if it can be computed by a deterministic TM in polynomial time.

A function $f \colon \{0, 1\}^* \rightarrow \N$ is in $\#\mathrm{P}$ if there exists a polynomial $p \colon \N \rightarrow \N$ and a polynomial time TM $M$, such that for every string $x\in \{0, 1\}^*$,
\begin{equation*}
	f(x)=|\{y\in\{0,1\}^{p(|x|)}\colon M(x,y)=1\}|.
\end{equation*}
\end{defn}

\begin{rem}
Definition \ref{def:FPsharpP} can also be described using NDTMs. $\#\mathrm{P}$ consist of all functions $f$, such that $f(x)$ equals the number of certificates that describes the sequence of moves of an accepting path of a polynomial time NDTM $M$ on input $x$.
\end{rem}
\begin{rem}
By Definition \ref{def:FPsharpP}, it is evident that $\mathrm{FP} \subseteq \#P$. However, similar to $\mathrm{P}$ vs. $\mathrm{NP}$ it is an open question whether $\mathrm{FP} = \#P$, i.e., whether any problem in $\#\mathrm{P}$ can be efficiently (in polynomial time) solved by a TM. It is commonly assumed that $\mathrm{FP} \varsubsetneq \#P$. Moreover, if $\mathrm{FP} = \#P$, then this would
imply that $P= NP$. Conversely, $P\neq NP$ implies $FP\neq \#P$.
\end{rem}

%
With this, we can argue that problems in $\#\mathrm{P}$ and $\#\mathrm{P}_1$ are considerably more complex than problems in $\mathrm{NP}$ and $\mathrm{NP}_1,$ respectively. In $\mathrm{NP}$, it is generally challenging to find a $y_*$ for $x$ such that
\begin{equation}\label{eq:c}
 f(x,y_*)=1
\end{equation}
assuming the commonly accepted complexity assumption that $P\neq NP$. 
If $\hat{y}$ is considered a potential solution for \eqref{eq:c}, it is a straightforward task to determine whether $f(x,\hat{y})=1$ in polynomial time. The verification of whether $\hat{y}$ is a solution for the problem or not can be performed in a simple manner. It should be noted that this asymmetry between finding the solution and verifying if a given value is a solution or not forms the foundation of cryptography as a whole.

For $\#\mathrm{P}$, such a behavior is unknown. Even if $\{y_1,\dots,y_r\}$ is considered as a solution set, the verification of whether $f(x,y_l)=1$ for $1\leq l\leq r$ is easy to implement. However, it is not clear if this approach is useful for $\#\mathrm{P}$ since there may exist additional solutions not included in the considered solution set. In the case of $\#\mathrm{P}$, we require the set of all possible solutions. These arguments also apply to the sets $\mathrm{NP}_1$ and $\#\mathrm{P}_1$.

For the next definition we need the notion of reduction. A reduction is an algorithm for transforming one problem into another problem.
\begin{defn}\label{def:sharpPcomplete} 
A function $f \in \#P$ is said to be complete in $\#\mathrm{P}$ if any other $g \in \#P$ can be reduced to $f$ by a polynomial time TM.
\end{defn}
\begin{rem}
Suppose there is a deterministic TM capable of solving a problem $f$ in polynomial time. In such a scenario, it implies that any other problem $g$ belonging to the complexity class $\#\mathrm{P}$ can also be solved in polynomial time by a deterministic TM. In simpler terms, if $f$ is considered as a complete problem within the class $\#\mathrm{P}$, then $f$ is at least as challenging as the most difficult problem in $\#\mathrm{P}$.
\end{rem}

 In this work, we are interested in studying functions that are defined on the singleton alphabet, i.e., $\{0\}^* \subset \{0, 1\}^*$. In other words, these functions are defined solely on the set of finite words composed of the symbol $0$. The classes analog to $\mathrm{FP}$ and $\#\mathrm{P}$ defined on singleton sets are denoted by $\mathrm{FP}_1$ and $\#\mathrm{P}_1$ respectively.

\begin{defn}[Classes $\mathrm{FP}_1$ and $\#\mathrm{P}_1$]\label{def:FP1sharpP1}
A function $f \colon\{0\}^* \to \N$ is said to be in $\mathrm{FP}_1$ if it can be computed by a deterministic TM in polynomial time.

A function $f\colon \{0\}^* \to \N$ is said to be in $\#\mathrm{P}_1$ if there exists a polynomial $p \colon \N \to \N$ and a polynomial-
time TM $M$ so that for every string $x \in \{0\}^*$
\begin{equation*}
	f(x)=|\{y\in\{0\}^{p(|x|)}\colon M(x,y)=1\}|.
\end{equation*} 
\end{defn}

\begin{rem}
As in the previous cases, we have $\mathrm{FP}_1 \subseteq \#\mathrm{P}_1$ but it is open whether $\mathrm{FP}_1 = \#\mathrm{P}_1$. However, it is widely assumed that $\mathrm{FP}_1 \varsubsetneq \#\mathrm{P}_1$. Similarly as above, an $f \in \#\mathrm{P}_1$ is said to be complete in $\#\mathrm{P}_1$ if any other $g\in \#\mathrm{P}_1$ can be reduced to $f$ by a polynomial time TM.
\end{rem}

\begin{rem}
Similar to the relation between the classes $\mathrm{NP}$ and $\#\mathrm{P}$, the class $\#\mathrm{P}_1$ is more general than $\mathrm{NP}_1$ since it not only contains problems for which a certificate can be verified in polynomial time but also counts the number of certificates verifiable in polynomial time. $\#\mathrm{P}_1$ is therefore more difficult and more complex than $\mathrm{NP}_1$.
\end{rem}

\subsection{Computation and Complexity of Integration}

The computational complexity study of the capacity of band-limited ACGN channels and their capacity-achieving p.s.d.s relies on the computation and computational complexity of integration. In this subsection we present the main results that we need for determining the computational complexity of the ACGN capacity and its capacity-achieving p.s.d.

We first need a result regarding the computation of the integral. This is introduced in the following theorem.

\begin{thm}[\cite{pour2017computability}]\label{thm:intcomp}
Let $[a,b]\subset\R$ be a compact interval. If $f\colon[a,b]\rightarrow\R$ is a computable function then $\int_a^bf(t)\,dt$ is a computable number.
\end{thm}

The first results regarding the computatoinal complexity of integrals were derived in \cite{friedman1984computational}. The following theorem states that the integral of a polynomial time computable function over an interval gives a polynomial time computable number if $\mathrm{FP}_1=\#\mathrm{P}_1$.

\begin{thm}[\cite{ko1986approximation}]\label{thm:intsharpP1compl}
The number $\int_a^b f(t)\,dt$ is polynomial time computable for all polynomial time computable functions $f\colon[a,b]\to\R$ if and only if $\mathrm{FP}_1=\#\mathrm{P}_1$.
\end{thm}

\section{Complexity Blowup of the ACGN Capacity Computation}
\label{sec:BlowupACGN}
In this section we focus on studying the computational complexity property of the capacity-achieving p.s.d. and the capacity of the band-limited ACGN channel. The goal is to answer Question 1 and Question 2.

First we focus on the capacity-achieving p.s.d.. The capacity-achieving p.s.d. of the band-limited ACGN, can be approached using the water pouring technique. To study the complexity of computing the capacity-achieving p.s.d. we consider a band-limited channel with rational bandwidth and power constraint and a polynomial time computable continuous noise spectral density. The following theorem classifies the complexity class to which the capacity-achieving p.s.d. belongs.

\begin{thm}\label{thm:1}
	Let $B\in\R_c$ be a polynomial time computable number, $N\colon [0,B]\rightarrow \R$ be a polynomial time computable continuous function and $P\in\Q$ with $P>\bar{N}B$ be arbitrary. Then $P_x^*(P,0)$ is in $\#\mathrm{P}_1$. Furthermore, there exists a strictly positive computable noise p.s.d. $N_*$ that is infinitely differentiable on $[0,B]$, such that for all $P>\bar{N}_*B$, where $\bar{N}_*$ is the average noise p.s.d. and $P\in\Q$, the function $P_x^*(P,0)$ is complete in $\#\mathrm{P}_1$.
\end{thm}

\begin{proof}
Let $N$ be a strictly positive and polynomial time computable noise p.s.d.. We have that 
\begin{equation*}
	P_x^*(P,f)=L-N(f), \quad f\in[0,B]
\end{equation*}
with
\begin{equation*}
L=\frac{1}{B}(P-\int_0^BN(f)\,df).
\end{equation*}
Since $B$ is a polynomial time computable number, we have that $\frac{1}{B}$ is also polynomial time computable. $P\in\Q$ and hence $\mathrm{P}$ is also a polynomial time computable number. From Theorem \ref{thm:intsharpP1compl}, we have that the computation of $\int_0^BN(f)\,df$ is in $\#\mathrm{P}_1$. This implies that $L$ is also in $\#\mathrm{P}_1$. This proves the first statement of the theorem.

From Theorem \ref{thm:intsharpP1compl}, we have that there exists a function $g\in\sC_c([0,1])$ that is polynomial time computable, such that the computation of $\int_0^1 g(f)\,df$ is $\#\mathrm{P}_1$-complete. 
We consider the function 

\begin{equation*}
g_1(f)\coloneqq \frac{1}{B} g\Big(\frac{f}{B}\Big) \quad 0\leq f\leq B.
\end{equation*}
$g_1$ is an infinitely differentiable function, and it is polynomial time computable. There is a rational number $\alpha$ such that for every $\omega\in[0,1]$
\begin{equation*}
	N_*(f)=\alpha+g_1(\omega)
\end{equation*}
fulfills the condition $\min_{f\in[0,B]}N_*(f)>0$.

Then, for $P\in\Q$, $P>\frac{1}{B}\int_0^BN_*(f)\,df=\bar{N}_*$ we have
\begin{equation*}
	L=\frac{1}{B}(P+\int_0^BN_*(f)\,df).
\end{equation*}
Thus with
\begin{equation*}
	\int_0^B N_*(f)\,df=\int_0^1 g(f)\,df+\alpha B
\end{equation*}
we have 
\begin{equation*}
	LB - P-\alpha B = \int_0^1N_*(f)\,df.
\end{equation*}
This way, the computation of the number $LB - P-\alpha B$ is complete in $\#\mathrm{P}_1$. Since $P, \alpha\in\Q$ and $B$ are polynomial time computable, we have that the computation of $L$ is $\#\mathrm{P}_1$-complete.

Now we have 
\begin{equation*}
	P_x^*(P,0)=L-N_*(0).
\end{equation*}
Since $N_*(0)$ is polynomial time computable, we have that the computation of $P_x^*(P,0)$ is $\#\mathrm{P}_1$-complete.
\end{proof}

\begin{rem}
Theorem \ref{thm:1} considers parameters with simple characteristics, specifically, rational bandwidth and power constraints, along with continuous noise spectral densities that are computable in polynomial time. Since both the bandwidth and power constraints are rational, their computation can also be achieved in polynomial time. While all three parameters are polynomial time computable, it is only established that the capacity-achieving p.s.d. belongs to the class $\#\mathrm{P}_1$. Whether the capacity-achieving p.s.d. can be expressed as a polynomial time computable function of the frequency remains an unanswered question. Resolving this question would require understanding the relationship between $\mathrm{FP}_1$ and $\#\mathrm{P}_1$, which is currently an open problem. It is widely assumed, though not proven, that $\mathrm{FP}_1 \neq \#\mathrm{P}_1$. If this assumption holds true, then we can derive the following result.
\end{rem}

\begin{rem}
From Theorem \ref{thm:1}, we have that computing the capacity-achieving p.s.d. of the band-limited ACGN channel is complete in $\#\mathrm{P}_1$. This problem is more complex than $\mathrm{NP}$-complete problems. $\#\mathrm{P}$-complete problems are generally more complex than $\mathrm{NP}$-complete problems. While both classes represent computationally challenging problems, $\#\mathrm{P}$-complete problems involve counting or enumerating solutions, which typically requires more computational resources than verifying a solution (as in $\mathrm{NP}$-complete problems).
\end{rem}

\begin{rem} 
Even computing the capacity-achieving distribution for discrete memoryless channels (DMCs) is challenging. While there exist algorithms that can compute the capacity of DMCs \cite{arimoto1972algorithm, blahut1972computation}, a general stopping criterion for computing the capacity-achieving distributions for DMCs cannot exist \cite{boche2022algorithmic}. Even in cases where the capacity-achieving distribution of DMCs becomes computable, no definitive assertions have been made about the complexity of computing the optimal distribution. On the other hand, in the case of continuous channels, more precisely, the band-limited ACGN channel, Theorem \ref{thm:1} demonstrates that computing the capacity-achieving p.s.d. is $\#\mathrm{P}_1$-complete, indicating its hardness comparable to solving any problem in $\#\mathrm{P}_1$.
\end{rem}

\begin{thm}\label{thm:2}
If $\mathrm{FP}_1\neq \#\mathrm{P}_1$, then there exists an infinitely differentiable noise p.s.d. $N_*$ in $[0,B]$ that is polynomial time computable and for which $P_x^*(P,0)$ cannot be computed in polynomial time.
\end{thm}

\begin{proof}
With Theorem \ref{thm:1}, we have the complete characterization of the computational complexity of the capacity-achieving p.s.d. for $f=0$ fully characterized. This result holds for all rational frequencies $f\in[0,B]$.
\end{proof}

\begin{rem}
Based on Theorem \ref{thm:2}, if the widely accepted assumption $\mathrm{FP}_1 \neq \#\mathrm{P}_1$ holds true, it implies the existence of a noise p.s.d. $N_*$ that can be computed in polynomial time. Remarkably, in this case, the capacity-achieving p.s.d. exhibits a complexity-blowup phenomenon.
\end{rem}

Next, we focus on the capacity function of the band-limited ACGN channel with the goal of answering Question 2. In \cite{boche2023algorithmic}, it was shown that there are infinitely many noise p.s.d.s whose capacity yields a non-computable number. In this work, we aim to characterize the classes of band-limited ACGN channels whose capacities yield a computable number. The following theorem provides a description of the structure of the noise p.s.d. that ensures the computability of the capacity.

\begin{thm}\label{thm:computableC}
If $N$ is a strictly positive and computable continuous noise p.s.d. and $P\in\Q$ with $P>\bar{N}B$, then the capacity $C(N,P)\in\R_c$.
\end{thm}
\begin{proof}
Let $N$ be a computable continuous function, such that $\min_{f\in[0,B]}N(f)>0$. From Eq.\eqref{eq:cL}, we have that 
\begin{equation}
	C(P,N)=B\ln L-\int_0^B\ln(N(f))\,df.
\end{equation}
The term $B\ln L$ is computable, since $B$ is computable, $L$ is computable and $\ln$ is a computable function. From the first statement of Lemma \ref{lem:6}, we have that $\ln N(f)$ is also a computable continuous function. And hence, from Theorem \ref{thm:intcomp}, we have that $\int_0^B\ln(N(f))\,df\in\R_c$.
\end{proof}

While we have characterized the class of ACGN channels for which the capacity becomes computable, our focus now shifts to studying the computational complexity involved in computing the capacity. In other words, we aim to determine the level of complexity associated with computing the capacity of band-limited ACGN channels, given a noise spectral density with low complexity and a power constraint that can be computed in polynomial time.

\begin{thm}\label{thm:3}
Let $B$ be a polynomial time computable number, and $N$ be a strictly positive and polynomial time computable noise p.s.d.. Then the computation of the capacity $C(P,N)$ for $P\in\Q$, $P>\bar{N}B$ is in $\#\mathrm{P}_1$.
Furthermore, there is an infinitely differentiable and strictly positive noise p.s.d. $N_*$ and a $P_*>\bar{N_*}B$ where $\bar{N_*}$ is the average noise p.s.d., such that the computation of $C(P_*,N_*)$ cannot be polynomial time computable if $\mathrm{FP}_1\neq\#\mathrm{P}_1$.
\end{thm}

\begin{proof}
Let $N$ be such that it satisfies the conditions of Theorem \ref{thm:3}.
Then from Lemma \ref{lem:6} we have that $\ln N$ is polynomial time computable, since $N$ is strictly a positive computable continuous function. Furthermore we have that $L>0$. Since the computation of $L$ is in $\#\mathrm{P}_1$ then from Lemma \ref{lem:6}
we have that the computation of $\ln L$ is also in $\#\mathrm{P}_1$.

	We start the proof by demonstrating the first statement. 
	For $P>\bar{N}B$, $P\in\Q$. From \eqref{eq:cL}, we have that the capacity has the following form
	\begin{equation*}
		C(P,N)= B\ln L - \int_0^B \ln N(f)\,df
	\end{equation*}
	We study the computation of each of the terms of the right hand side of \eqref{eq:cL}. In the first term of \eqref{eq:cL}, we have that $L>0$ and since the computation of $L$ is in $\#\mathrm{P}_1$, then the computation of $\ln L$ is also in $\#\mathrm{P}_1$.
	Now we consider the second term of \eqref{eq:cL}. Let $N$ be a noise p.s.d. satisfying the conditions of Theorem \ref{thm:3}. Then we have that $\ln N$ is also polynomial time computable, since $N$ is a strictly positive computable continuous function that is polynomial time computable. This implies that the computation of $C(P,N)$ is in $\#\mathrm{P}_1$. This way we have shown the first statement.
	
Now we prove the second statement. Let $\beta\in\Q$, $0<\beta \leq 1 $. We consider the following noise p.s.d. 
\begin{equation*}
	N(f,\beta)=\beta N_*(f).
\end{equation*}

For $P> B\bar{N}_*$, $\beta_1,\beta_2\in\Q$, $0<\beta_l \leq 1$, $l=1,2$ and $\beta_1\neq\beta_2$, we have that the corresponding capacity is
\begin{equation*}
	C(P,N(\cdot,\beta_l))=B\ln L_l-\int_0^B\ln N(f,\beta)\,df
\end{equation*}
with
\begin{align*}
L_l&=\frac{P}{B}+\frac{1}{B}\int_0^B\ln N(f,\beta_l)\,df= \frac{P}{B}+\frac{\beta_l}{B}\int_0^B\ln N_*(f)\,df.
\end{align*}
Assume that the computation of both numbers $C(P,N(\cdot,\beta_1))$ and $C(P,N(\cdot,\beta_2))$ are possible in polynomial time, then the computation of the number 
\begin{align*}
	C(P,N(\cdot,\mathrm{P}_1))&-C(P,N(\cdot,P_2))\\
	&= B\ln L_1-B\ln L_2-\beta_1B +\beta_2B\\
	&=B\ln\frac{L_1}{L_2}-B(\beta_1-\beta_2)
\end{align*}
is polynomial time computable. Since $\beta_1,\beta_2\in\Q$ we have that the number $B(\beta_1-\beta_2)$ is polynomial time computable as well. This way we have that the number $z=B\ln\frac{L_1}{L_2}$ is polynomial time computable. Hence, $\frac{L_1}{L_2}=2^{\frac{z}{B}}=c$ is polynomial time computable. 
We have that 
\begin{equation*}
	c = \frac{L_1}{ L_2}=\frac{P+\beta_1\int_0^B\ln N_*(f)\,df}{P+\beta_2\int_0^B\ln N_*(f)\,df},
\end{equation*}
and hence 
\begin{equation*}
	z_* = \int_0^B\ln N_*(f)\,df=\frac{P(c-1)}{\beta_1-c\beta_2},
\end{equation*}
where $\beta_1\neq c\beta_2$ and $c\neq 1$. Since $P,\beta_1,\beta_2\in\Q$ and $c$ are polynomial time computable, then $z_*$ must also be polynomial time computable. However, if $\mathrm{FP}_1\neq\#\mathrm{P}_1$, then this $z_*$ cannot be polynomial time computable.
This way we prove the second statement. That is, for $l=1,2$ at least one of the numbers $C(P,N(\cdot,\beta_l))$ is not polynomial time computable.
\end{proof}

\begin{rem}
Considering Theorem \ref{thm:3}, and assuming the widely accepted assumption $\mathrm{FP}_1 \neq \#\mathrm{P}_1$ is correct, we find that there exists a noise p.s.d. $N_*$ and a corresponding power constraint $P_*$ that can both be computed in polynomial time. Remarkably, under these conditions, the capacity of the band-limited ACGN channel demonstrates a complexity-blowup phenomenon.
\end{rem}

\begin{rem}
	Theorem \ref{thm:3} has also interesting consequences for coding theory. Assume that a sequence, as a function of the blocklength, of capacity-achieving codes can be found such that the encoding and decoding processes corresponding to the blocklength have polynomial complexity, then it is possible to calculate the code's rate as a function of the blocklength in polynomial complexity. 
	
If we intend to calculate the blocklength such that the capacity approximation $C(P_*,N_*)-\frac{1}{2^N}$ can be achieved, then the corresponding minimum blocklength grows as a function of the approximation parameter $N$ faster than any polynomial. Otherwise, $C(P_*,N_*)$ would be computable in polynomial time.
\end{rem}

\begin{rem}
Generally, the complexity problems that are studied aim to classify problems into the $\mathrm{P}$ or $\mathrm{NP}$ complexity classes. However, when it comes to computing the capacity-achieving p.s.d., which is $\#\mathrm{P}_1$-complete, the task is typically more challenging compared to solving an $\mathrm{NP}_1$-complete problem. Therefore computing the capacity-achieving p.s.d. and the capacity of the band-limited ACGN channels are harder problems than $\mathrm{NP}_1$.
\end{rem}

\section{Discussion} 
\label{sec:conclusion}

While the capacity of band-limited ACGN channels is generally not computable, we have successfully characterized the subset of these channels that do have computable capacities. Our findings demonstrate that as long as the continuous noise spectral densities are strictly positive and computable, the resulting capacity will always be a computable number.

Furthermore, we have studied the computational complexity involved in determining the capacity of such channels. Our analysis reveals that calculating the capacity of polynomial time computable continuous ACGN channels not only falls within the $\#\mathrm{P}_1$ class. 
Additionally, we have shown that if the widely accepted assumption $\mathrm{FP}_1\neq\#\mathrm{P}_1$ holds true, then it is impossible to compute the capacity of band-limited ACGN channels in polynomial time. 

This has also interesting consequences for coding theory. Assume that a sequence of capacity achieving codes has polynomial complexity, then computing code's rate as a function of the blocklength has also polynomial complexity. However the computation of the minimum blocklength corresponding to the capacity approximation as a function of the approximation error $N$ grows faster than any polynomial.

Moreover, we have explored the computational complexity of determining the capacity-achieving p.s.d. of band-limited ACGN channels. Our analysis examines the relationship between the complexity of computing the capacity-achieving p.s.d. and the computational complexity of the power constraint and the ACGN channel itself. Our results demonstrate that when considering polynomial time computable parameters, such as the power constraint, bandwidth, and noise p.s.d., the computation of the capacity-achieving p.s.d. becomes $\#\mathrm{P}_1$-complete.

This finding implies that if the assumption $\mathrm{FP}_1\neq\#\mathrm{P}_1$ holds true, there can exist a noise p.s.d. that is computable in polynomial time, while the capacity-achieving p.s.d. associated with it cannot be calculated within polynomial time. This demonstrates a complexity blow-up behavior, illustrating the increased difficulty of determining the capacity-achieving p.s.d. in such scenarios.

The focus of studying the complexity of problems in information and communication theory lies in determining whether a computational problem belongs to the classes $\mathrm{P}$ or $\mathrm{NP}$ (or $\mathrm{P}_1$ and $\mathrm{NP}_1$, respectively). In this paper, we show that both the computation of the capacity of band-limited ACGN channels and the computation of the capacity-achieving p.s.d. fall within the class $\#\mathrm{P}_1$. In $\mathrm{NP}$, one can efficiently verify a single certificate for a problem. However, $\mathrm{NP}_1$ does not provide information about the number of certificates. $\#\mathrm{P}_1$ counts the certificates that can be efficiently verified, making it a more general and complex class than $\mathrm{NP}_1$. And hence problems classified as $\#\mathrm{P}_1$-complete are harder than problems classified as $\mathrm{NP}_1$-complete, under the common complexity assumptions.


\appendix
In this appendix, we introduce lemmas describing the algorithmic properties and computational complexity of the logarithmic function, which are used to prove the results of this work.
 
\begin{lem}\label{lem:3}
Let $\underline{\alpha},\bar{\alpha}\in\Q$ with $0<\underline{\alpha}<\bar{\alpha}<\infty$. Let $x_*=\frac{\underline{\alpha}+\bar{\alpha}}{2}$ and $\beta=\frac{2\underline{\alpha}}{\bar{\alpha}+\underline{\alpha}}$. Let $\frac{r}{2^s}$ be a dyadic number with $s,r\in\N$ such that $\beta\leq \frac{r}{2^s}<1$ and let $m_1\in\N$ be such that $\frac{r^{m_1}}{2^{sm_1}}<\frac{1}{2}$. Then for all $x\in [\underline{\alpha},\bar{\alpha}]$ and for all $m \geq m_1$ we have
\begin{equation*}
	\Big|\ln x-\ln(x_*)-\sum_{\ell=1}^{m^2}\frac{(-1)^{\ell-1}}{\ell x_*^\ell}(x-x_*)^\ell\Big|<\gamma\frac{1}{2^m} 
\end{equation*}
with $\gamma=\frac{2\underline{\alpha}}{\bar{\alpha}-\underline{\alpha}}\in\Q$.
\end{lem}

\begin{proof}
	$\ln(\cdot)$ is an absolute convergent Taylor-series in $x\in[\underline{\alpha},\bar{\alpha}]$. For $\Psi(x)=\ln(x)$ we have $\Psi^{(\ell)}(x)=\frac{(-1)^{\ell-1}}{\ell x_*^\ell}$, for $\ell\geq 1$ and $x>0$.
	
We then have 
	\begin{align*}
	\Big|\ln x&-\ln(x_*)-\sum_{\ell=1}^{m^2}\frac{(-1)^{\ell-1}}{\ell x_*^\ell}(x-x_*)^\ell\Big|\\
	&= \Big|\sum_{\ell=m^2+1}^\infty\frac{(-1)^{\ell-1}}{\ell x_*^\ell}(x-x_*)^\ell\Big|
	\leq \sum_{\ell=m^2+1}^\infty\frac{|x-x_*|^\ell}{|x_*|^\ell}\\
	&= \sum_{\ell=m^2+1}^\infty\Big|1-\frac{x}{x_*}\Big|^\ell\leq\sum_{\ell=m^2+1}^\infty\Big|\frac{\underline{\alpha}+\bar{\alpha}-(\bar{\alpha}-\underline{\alpha})}{\underline{\alpha}+\bar{\alpha}}\Big|^\ell\\
	&=\sum_{\ell=m^2+1}^\infty\Big(\frac{2\underline{\alpha}}{\underline{\alpha}+\bar{\alpha}}\Big)^\ell
	=\sum_{\ell=m^2+1}^\infty\beta^\ell\\
	&=\beta^{m^2+1}\frac{1}{1-\beta}=\frac{\beta}{1-\beta}\beta^{M^2}\\
	&=\frac{\frac{2\underline{\alpha}}{\underline{\alpha}+\bar{\alpha}}}{1-\frac{2\underline{\alpha}}{\underline{\alpha}+\bar{\alpha}}}\Big(\frac{2\underline{\alpha}}{\underline{\alpha}+\bar{\alpha}}\Big)^{m^2}\\
	&=\frac{2\underline{\alpha}}{\bar{\alpha}-\underline{\alpha}}\Big(\frac{2\underline{\alpha}}{\underline{\alpha}+\bar{\alpha}}\Big)^{m^2}\\
	&\leq \frac{2\underline{\alpha}}{\bar{\alpha}-\underline{\alpha}}\frac{r^{m^2}}{2^{sm^2}}
	=\gamma \Big(\frac{r^m}{2^{sm}}\Big)^m
	<\gamma\frac{1}{2^m}.
	\end{align*}
\end{proof}
The constant $\gamma$, does not influence the binary convergence of the power series. So the power series binary converges to the logarithm function. This is visualized in the next lemma.

\begin{lem}\label{lem:4}
Let $m_1,m_2\in\N$ be arbitrary such that $m_2\geq m_1$. Let $\gamma<2^{m_2}$. Then for all $m\in\N$ and all $x$ as in Lemma \ref{lem:3} we have
	\begin{equation*}
		\Big|\ln x-\ln(x_*)-\sum_{\ell=1}^{(m_2+m)^2}\frac{(-1)^{\ell-1}}{\ell x_*^\ell}(x-x_*)^{\ell}\Big|<\frac{1}{2^m}.
	\end{equation*}
\end{lem}

\begin{proof}
The proof follows from Lemma \ref{lem:3}. We have
	\begin{equation*}
		\Big|\sum_{\ell=(m_2+m)^2}^{\infty}\frac{(-1)^{\ell-1}}{\ell x_*^\ell}(x-x_*)^{\ell}\Big|<\gamma \frac{1}{2^{m_2+m}}<\frac{1}{2^m}.
	\end{equation*}
\end{proof}

\begin{lem}\label{lem:5}
 For $x\in[\underline{\alpha},\bar{\alpha}]$, $\beta=\frac{\bar{\alpha}-\underline{\alpha}}{\bar{\alpha}+\underline{\alpha}}$ and $m\in\N$ we consider the polynomial
 \begin{equation*}
 Q_m(x)=\ln(x_*)+\sum_{\ell=1}^{m^2}\frac{(-1)^{\ell-1}}{\ell x_*^\ell}(x-x_*)^\ell.
 \end{equation*}
 For $x_1,x_2\in[\underline{\alpha},\bar{\alpha}]$ we have
 \begin{equation*}
 |Q_m(x_1)-Q_m(x_2)|\leq \frac{2}{\beta(1-\beta)^2}|x_1-x_2|.
 \end{equation*}
\end{lem}

\begin{proof}
	Using the mean value theorem we get for  $f_\ell(x)=(x-x_*)^{\ell}$ for $1\leq\ell\leq m^2$
	\begin{align*}
	 |f_\ell(x_1)-f_\ell(x_2)|&=|f'_\ell (x_{1,2})||x_1-x_2|\\
	 &\leq \ell|(\bar{\alpha}-x_*)^{\ell-1}||x_1-x_2|\\
	 &=\ell|(\frac{\bar{\alpha}-\underline{\alpha}}{2})^{\ell-1}||x_1-x_2|.
	\end{align*}
	We then have that 
	\begin{align*}
		|Q_m(x_1)-Q_m(x_2)|&\leq \sum_{\ell=1}^{m^2}\frac{(-1)^{\ell+1}}{\ell x_*^\ell}\ell(\frac{\bar{\alpha}-\underline{\alpha}}{2})^{\ell-1}||x_1-x_2|\\
		&\leq \frac{|x_1-x_2|}{x_*}\sum_{\ell=1}^{m^2}\Big(\frac{\frac{\bar{\alpha}-\underline{\alpha}}{2}}{\frac{\bar{\alpha}+\underline{\alpha}}{2}}\Big)^{\ell-1}\\
		&< \frac{|x_1-x_2|}{x_*}\sum_{\ell=1}^{\infty}\beta^{\ell-1}\\
		&\leq \frac{|x_1-x_2|}{x_*}\sum_{\ell=0}^{\infty}\beta^{\ell}\\
		&=\frac{|x_1-x_2|}{x_*}\frac{1}{1-\beta}\\
		&=\frac{2}{\beta(1-\beta)^2}|x_1-x_2|.
	\end{align*}
\end{proof}

Next we look at the computability and complexity properties of the logarithm function. More precisely, we show that the composition of a computable continuous periodic function $g$ with the logarithm function, i.e. $\ln \circ g(\omega)$, results in a computable continuous function. Moreover, if the function $g$ has low complexity, then the composition $\ln \circ g(\omega)$ has also low complexity.

\begin{lem}\label{lem:6}
Let $g$ be a computable continuous $2\pi$-periodic function with $\min_{\omega\in[-\pi,\pi]}g(\omega)=\underline{c}>0$. 
\begin{enumerate}
	\item Then $\ln g$ is also a computable continuous $2\pi$-periodic function.
	\item Let $g$ polynomial time computable, then $\ln g$ is also polynomial time computable.
\end{enumerate}
\end{lem}

\begin{proof}
Let $g$ be a fixed function. Let 
\begin{equation*}
	\underline{c} = \min_{\omega\in[-\pi,\pi]}g(\omega)>0
\text{ and } 
	\bar{C}=\max_{\omega\in[-\pi,\pi]}g(\omega)<\infty.
\end{equation*}
We choose a $\underline{\alpha}\in\Q$ with $\underline{\alpha}<\frac{\underline{c}}{2}$ and $\bar{\alpha}\in\Q$ with $\bar{\alpha}>\bar{C}+1$. 
In this proof, we work with the oracle Turing model. We start with an algorithm for computing $g$. Based on this algorithm, we construct a new algorithm for computing $\ln g$.
If the algorithm computing $g$ computes the approximation of $g$ with an approximation error of $\frac{1}{2^M}$ in polynomial time, then the new constructed algorithm for $\ln g$ also computes an approximation of $\ln g$ with an approximation error of $\frac{1}{2^M}$ in polynomial time. 

We choose $m_3\in\N$ such that $\frac{2}{\beta(1-\beta)^2}<2^{m_3}$ holds.
Let $M\in\N$ be arbitrary. Let $C(\omega,M)$ be the approximation of $g$ with precision $M$, i.e.,
\begin{equation*}
|g(\omega)-C(\omega,M)|<\frac{1}{2^M}.
\end{equation*}

Let the number $C(\omega,\tilde{M})$ with $\tilde{M}=M+m_3+1+m_2$ be the approximation for $g(\omega)$ with approximation error $\frac{1}{2^M}$ computed by the algorithm to calculate $g(\omega)$ for input $M$ and Oracle input $\omega$. We then have
\begin{equation*}
	|g(\omega) - C(\omega,\tilde{M})|<\frac{1}{2^{M+1}}.
\end{equation*}
We use the number $C(\omega,\tilde{M})$ and calculate $d(\omega,M)\coloneqq Q_{\tilde{M}}(C(\omega,\tilde{M}))$.

The polynomial 
	\begin{equation*}
		Q_{\tilde{M}}(x)=\sum_{\ell=0}^{M^2}\frac{(-1)^{\ell+1}}{x_*^\ell}(x-x_*)^\ell.
	\end{equation*}
has only polynomial many coefficients that are different from $0$. All of them are rational numbers and are computed depending on the precision $\tilde{M}$ in polynomial time. This way, the approximation $d(\omega,M)$ of the number $C(\omega,\tilde{M})$ can be computed in polynomial time depending on $M$. 
Now we have
\begin{align*}
|\ln g(\omega)-d(\omega,M)|&=|\ln g(\omega)-Q_{\tilde{M}}(g(\omega))\\
&\quad +Q_{\tilde{M}}(g(\omega))-Q_{\tilde{M}}(C(\omega,\tilde{M}))|\\
&\leq |\ln g(\omega)-Q_{\tilde{M}}(g(\omega))|\\
&\quad+|Q_{\tilde{M}}(g(\omega))-Q_{\tilde{M}}(C(\omega,\tilde{M}))|.
\end{align*}
From Lemma \ref{lem:4} we have that 
\begin{equation*}
	|\ln g(\omega)-Q_{\tilde{M}}(g(\omega))|<\frac{1}{2^{M+1}}.
\end{equation*}
Further, from Lemma \ref{lem:5} and the definition of $m_3$ we have
\begin{align*}
	|Q_{\tilde{M}}(g(\omega))-Q_{\tilde{M}}(C(\omega,\tilde{M}))|&\leq|g(\omega)-C(\omega,\tilde{M})|<\frac{1}{2^{M+1}}.
\end{align*}
This way we have 
\begin{equation*}
	|\ln g(\omega)-d(\omega,M)|<\frac{1}{2^{M+1}}+\frac{1}{2^{M+1}}.
\end{equation*}
\end{proof}

\balance
\bibliographystyle{IEEEtran}
\bibliography{IEEEabrv,confs-jrnls,references_coloregaussian}

\begin{thebibliography}{10}
\providecommand{\url}[1]{#1}
\csname url@samestyle\endcsname
\providecommand{\newblock}{\relax}
\providecommand{\bibinfo}[2]{#2}
\providecommand{\BIBentrySTDinterwordspacing}{\spaceskip=0pt\relax}
\providecommand{\BIBentryALTinterwordstretchfactor}{4}
\providecommand{\BIBentryALTinterwordspacing}{\spaceskip=\fontdimen2\font plus
\BIBentryALTinterwordstretchfactor\fontdimen3\font minus
  \fontdimen4\font\relax}
\providecommand{\BIBforeignlanguage}[2]{{%
\expandafter\ifx\csname l@#1\endcsname\relax
\typeout{** WARNING: IEEEtran.bst: No hyphenation pattern has been}%
\typeout{** loaded for the language `#1'. Using the pattern for}%
\typeout{** the default language instead.}%
\else
\language=\csname l@#1\endcsname
\fi
#2}}
\providecommand{\BIBdecl}{\relax}
\BIBdecl

\bibitem{shannon1949communication}
C.~E. Shannon, ``Communication theory of secrecy systems,'' \emph{Bell Syst.
  Tech.~J.}, vol.~28, no.~4, pp. 656--715, Oct. 1949.

\bibitem{shannon1949communicationnoise}
------, ``Communication in the presence of noise,'' \emph{Proc. IRE}, vol.~37,
  no.~1, pp. 10--21, 1949.

\bibitem{wyner1966capacity}
A.~D. Wyner, ``The capacity of the band-limited {G}aussian channel,''
  \emph{Bell Syst. Tech.~J.}, vol.~45, no.~3, pp. 359--395, 1966.

\bibitem{ash1963capacity}
R.~B. Ash, ``Capacity and error bounds for a time-continuous {G}aussian
  channel,'' \emph{Inf. Contr.}, vol.~6, no.~1, pp. 14--27, 1963.

\bibitem{shannon1959probability}
C.~E. Shannon, ``Probability of error for optimal codes in a {G}aussian
  channel,'' \emph{Bell Syst. Tech.~J.}, vol.~38, no.~3, pp. 611--656, 1959.

\bibitem{ash2012information}
R.~B. Ash, \emph{Information {T}heory}.\hskip 1em plus 0.5em minus 0.4em\relax
  Courier Corporation, 2012.

\bibitem{cover1999elements}
T.~M. Cover, \emph{Elements of {I}nformation {T}heory}.\hskip 1em plus 0.5em
  minus 0.4em\relax John Wiley \& Sons, 1999.

\bibitem{ihara1993information}
S.~Ihara, \emph{Information {T}heory for {C}ontinuous {S}ystems}.\hskip 1em
  plus 0.5em minus 0.4em\relax World Scientific, 1993, vol.~2.

\bibitem{tse2005fundamentals}
D.~Tse and P.~Viswanath, \emph{Fundamentals of {W}ireless
  {C}ommunication}.\hskip 1em plus 0.5em minus 0.4em\relax Cambridge University
  Press, 2005.

\bibitem{pinsker1964information}
M.~S. Pinsker, \emph{Information and {I}nformation {S}tability of {R}andom
  {V}ariables and {P}rocesses}.\hskip 1em plus 0.5em minus 0.4em\relax
  Holden-Day, 1964.

\bibitem{gallager1968information}
R.~G. Gallager, \emph{Information Theory and Reliable Communication}.\hskip 1em
  plus 0.5em minus 0.4em\relax John Wiley \& Sons, Inc., 1968.

\bibitem{forney1998modulation}
G.~Forney and G.~Ungerboeck, ``Modulation and coding for linear {G}aussian
  channels,'' \emph{{IEEE} Trans. Inf. Theory}, vol.~44, no.~6, pp. 2384--2415,
  Oct. 1998.

\bibitem{turing1936computable}
A.~M. Turing \emph{et~al.}, ``On computable numbers, with an application to the
  {E}ntscheidungsproblem,'' \emph{Proc. London Math. Soc.}, vol.~2, no.~42, pp.
  230--265, 1936.

\bibitem{turing1938computable}
A.~M. Turing, ``On computable numbers, with an application to the
  {E}ntscheidungsproblem. {A} correction,'' \emph{Proc. London Math. Soc.},
  vol.~2, no.~43, pp. 544--546, 1937.

\bibitem{weihrauch2000computable}
K.~Weihrauch, \emph{Computable {A}nalysis: {A}n {I}ntroduction}.\hskip 1em plus
  0.5em minus 0.4em\relax Springer Science \& Business Media, 2000.

\bibitem{boche2023algorithmic}
H.~Boche, A.~Grigorescu, R.~F. Schaefer, and H.~V. Poor, ``Algorithmic
  computability of the capacity of {G}aussian channels with colored noise,''
  \emph{arXiv preprint arXiv:2305.02819}, 2023.

\bibitem{beigi2007complexity}
S.~Beigi and P.~W. Shor, ``On the complexity of computing zero-error and
  {H}olevo capacity of quantum channels,'' \emph{arXiv preprint
  arXiv:0709.2090}, 2007.

\bibitem{shannon1956zero}
C.~Shannon, ``The zero error capacity of a noisy channel,'' \emph{{IRE} Trans.
  Inf. Theory}, vol.~2, no.~3, pp. 8--19, Sep. 1956.

\bibitem{sipser1996introduction}
M.~Sipser, ``Introduction to the theory of computation,'' \emph{ACM Sigact
  News}, vol.~27, no.~1, pp. 27--29, 1996.

\bibitem{langberg2011hardness}
M.~Langberg and A.~Sprintson, ``On the hardness of approximating the network
  coding capacity,'' \emph{{IEEE} Trans. Inf. Theory}, vol.~57, no.~2, pp.
  1008--1014, Jan. 2011.

\bibitem{berlekamp1978inherent}
E.~Berlekamp, R.~McEliece, and H.~Van~Tilborg, ``On the inherent intractability
  of certain coding problems (corresp.),'' \emph{{IEEE} Trans. Inf. Theory},
  vol.~24, no.~3, pp. 384--386, May 1978.

\bibitem{vardy1997intractability}
A.~Vardy, ``The intractability of computing the minimum distance of a code,''
  \emph{{IEEE} Trans. Inf. Theory}, vol.~43, no.~6, pp. 1757--1766, Nov. 1997.

\bibitem{vardy1997algorithmic}
------, ``Algorithmic complexity in coding theory and the minimum distance
  problem,'' in \emph{Proc. ACM Symp. Theory of Comp.}, May 1997, pp. 92--109.

\bibitem{gaborit2016hardness}
P.~Gaborit and G.~Z{\'e}mor, ``On the hardness of the decoding and the minimum
  distance problems for rank codes,'' \emph{{IEEE} Trans. Inf. Theory},
  vol.~62, no.~12, pp. 7245--7252, Oct. 2016.

\bibitem{ordentlich2011two}
E.~Ordentlich and R.~M. Roth, ``Two-dimensional maximum-likelihood sequence
  detection is {NP} hard,'' \emph{{IEEE} Trans. Inf. Theory}, vol.~57, no.~12,
  pp. 7661--7670, Dec. 2011.

\bibitem{leditzky2020playing}
F.~Leditzky, M.~A. Alhejji, J.~Levin, and G.~Smith, ``Playing games with
  multiple access channels,'' \emph{Nature Commun.}, vol.~11, no.~1, p. 1497,
  Mar. 2020.

\bibitem{calvo2010computation}
E.~Calvo, D.~P. Palomar, J.~R. Fonollosa, and J.~Vidal, ``On the computation of
  the capacity region of the discrete {MAC},'' \emph{{IEEE} Trans. Inf.
  Theory}, vol.~58, no.~12, pp. 3512--3525, Sep. 2010.

\bibitem{mondelli2014polar}
M.~Mondelli, S.~H. Hassani, and R.~L. Urbanke, ``From polar to {R}eed-{M}uller
  codes: A technique to improve the finite-length performance,'' \emph{{IEEE}
  Trans. Commun.}, vol.~62, no.~9, pp. 3084--3091, Aug. 2014.

\bibitem{eslami2013finite}
A.~Eslami and H.~Pishro-Nik, ``On finite-length performance of polar codes:
  stopping sets, error floor, and concatenated design,'' \emph{{IEEE} Trans.
  Commun.}, vol.~61, no.~3, pp. 919--929, Feb. 2013.

\bibitem{sun2018distributed}
Y.~Sun, M.~Peng, and H.~V. Poor, ``A distributed approach to improving spectral
  efficiency in uplink device-to-device-enabled cloud radio access networks,''
  \emph{{IEEE} Trans. Commun.}, vol.~66, no.~12, pp. 6511--6526, Jul. 2018.

\bibitem{bagaa2017optimal}
M.~Bagaa, A.~Chelli, D.~Djenouri, T.~Taleb, I.~Balasingham, and K.~Kansanen,
  ``Optimal placement of relay nodes over limited positions in wireless sensor
  networks,'' \emph{{IEEE} Trans. Wireless Commun.}, vol.~16, no.~4, pp.
  2205--2219, Jan. 2017.

\bibitem{chattopadhyay2017gibbsian}
A.~Chattopadhyay, B.~B{\l}aszczyszyn, and H.~P. Keeler, ``Gibbsian on-line
  distributed content caching strategy for cellular networks,'' \emph{{IEEE}
  Trans. Commun.}, vol.~17, no.~2, pp. 969--981, Nov. 2017.

\bibitem{zhang2019cellular}
S.~Zhang, H.~Zhang, B.~Di, and L.~Song, ``Cellular {UAV}-to-{X} communications:
  Design and optimization for multi-{UAV} networks,'' \emph{{IEEE} Trans.
  Commun.}, vol.~18, no.~2, pp. 1346--1359, Jan. 2019.

\bibitem{flanagan2006proving}
M.~F. Flanagan, ``On proving the water pouring theorem for information rate
  optimization,'' in \emph{Int. Conf. on Signals and Electron. Syst.}\hskip 1em
  plus 0.5em minus 0.4em\relax Citeseer, 2006.

\bibitem{pour2017computability}
M.~B. Pour-El and J.~I. Richards, \emph{Computability in {A}nalysis and
  {P}hysics}.\hskip 1em plus 0.5em minus 0.4em\relax Cambridge University
  Press, 2017.

\bibitem{friedman1984computational}
H.~Friedman, ``The computational complexity of maximization and integration,''
  \emph{Advances in Mathematics}, vol.~53, no.~1, pp. 80--98, 1984.

\bibitem{ko1986approximation}
K.-I. Ko, ``Approximation to measurable functions and its relation to
  probabilistic computation,'' \emph{Annals of Pure and Appl. Logic}, vol.~30,
  no.~2, pp. 173--200, 1986.

\bibitem{arimoto1972algorithm}
S.~Arimoto, ``An algorithm for computing the capacity of arbitrary discrete
  memoryless channels,'' \emph{{IEEE} Trans. Inf. Theory}, vol.~18, no.~1, pp.
  14--20, Jan. 1972.

\bibitem{blahut1972computation}
R.~Blahut, ``Computation of channel capacity and rate-distortion functions,''
  \emph{{IEEE} Trans. Inf. Theory}, vol.~18, no.~4, pp. 460--473, Jul. 1972.

\bibitem{boche2022algorithmic}
H.~Boche, R.~F. Schaefer, and H.~V. Poor, ``Algorithmic computability and
  approximability of capacity-achieving input distributions,'' (in press).

\end{thebibliography}
\end{document}